\documentstyle[12pt]{article}

\newcommand{\ie}{{\textit i}.{\textit e}.}

\newcommand{\Nc}{N_{{\rm c}}}
\newcommand{\Nf}{N_{{\rm f}}}
\renewcommand{\L}{{\rm L}}
\newcommand{\R}{{\rm R}}
\newcommand{\GSM}{G_{{\rm SM}}}
\newcommand{\GSC}{G_{{\rm SC}}}

\newcommand{\bm}[1]{\mbox{\boldmath$#1$}}
\newcommand{\fun}[1]{\!\left(#1\right)}
\newcommand{\wb}[1]{%
\vbox{\ialign{##\crcr\hskip 1.0pt\hrulefill\hskip 0.3pt%
\crcr\noalign{\kern-1pt\vskip0.07cm\nointerlineskip}%
$\hfil\displaystyle{#1}\hfil$\crcr}}}
\begin{document}
\begin{titlepage}
\vspace*{-1.5cm}
  \begin{flushright}
    KANAZAWA-02-10\\[-1mm]
    KUNS-1781\\[-1mm]
    NIIG-DP-02-04\\[-1mm]
  \end{flushright}
  \begin{center}
    
    {\Large\bf 
    Yukawa Hierarchy Transfer \\[-2pt]
    Based on Superconformal Dynamics \\[6pt]
    and Geometrical Realization in String Models
    } 

\vspace*{1cm}
    Tatsuo~{\sc Kobayashi}\rlap,\,\footnote{
    E-mail: kobayash@gauge.scphys.kyoto-u.ac.jp}
    Hiroaki~{\sc Nakano}\rlap,\,\footnote{
    E-mail: nakano@muse.hep.sc.niigata-u.ac.jp}
    Tatsuya~{\sc Noguchi}\rlap,\,\footnote{
    E-mail: noguchi@gauge.scphys.kyoto-u.ac.jp}
and
    Haruhiko~{\sc Terao}\,\footnote{
    E-mail: terao@hep.s.kanazawa-u.ac.jp}\\

\vspace*{5mm}
  
  $^{1,3}$\textit{Department of Physics, 
Kyoto University, Kyoto 606-8502, Japan}\\
  $^{2}$\textit{Department of Physics, Niigata University, 
Niigata 950-2181, Japan}\\
  $^{4}$\textit{Institute for Theoretical Physics, Kanazawa
    University, Kanazawa 920-1192, Japan}\\
  \vspace{0.5cm}
  
\begin{abstract}
We propose a scenario that leads to hierarchical Yukawa couplings 
and degenerate sfermion masses at the same time,
in the context of extra-dimensional models, 
which can be naturally embedded in a wide class of string models.
The hierarchy of Yukawa couplings and degeneracy of 
sfermion masses can be realized 
thanks to superconformal gauge dynamics.
The sfermion mass degeneracy is guaranteed by taking the
superconformal fixed point to be family independent.
In our scenario, the origin of Yukawa hierarchy 
is attributed to geometry of compactified dimensions 
and the consequent volume dependence of gauge couplings
in the superconformal sectors.
The difference in these gauge couplings is dynamically 
transferred to the hierarchy of the Yukawa couplings.
Thus, our scenario combines 
a new dynamical approach and the conventional geometrical 
approach to the supersymmetric flavor problem.
\end{abstract}
\end{center}
\end{titlepage}

It is a great challenge to derive the realistic fermion masses 
and their mixing angles from superstring models.
Actually, one needs to explain hierarchical Yukawa couplings 
of quarks and leptons to electroweak Higgs fields,
\begin{equation}
y_{ij}\psi_i \psi_j H \ ,
\label{yukawa}
\end{equation}
where $\psi_i$ ($i=1,2,3$) represent three families of 
quarks and leptons collectively.
Here we concentrate on models with 
softly-broken ${\cal N}=1$ $D=4$ supersymmetry (SUSY), 
and $H$ denotes the up and down sectors of the Higgs fields, 
again collectively.
Several mechanisms for generating hierarchical Yukawa couplings
have been studied in the context of compactified string theories 
as well as four-dimensional effective field theories.

On the other hand,
when one supposes that the SUSY survives to low-energy world,
one should pay attention to the SUSY flavor problem;
in particular if sfermion masses are not sufficiently degenerate,
there appear unacceptably large flavor violations.
One approach to avoid such large flavor violations is 
to assume that SUSY breaking sector is completely sequesterred
from the Standard Model (SM) sector \cite{sequester}
and to find a flavor-blind mediation mechanism of SUSY breaking.
An alternative approach is to suppose some nontrivial flavor dynamics
that makes sfermions degenerate.
It is this latter approach that we pursue here.

In this letter, we propose a new scenario that simultaneously 
realizes Yukawa hierarchy and sfermion mass degeneracy,
independently of the origin of SUSY breaking. 
Our scenario is inspired by the work by Nelson and Strassler 
\cite{NS1}, in which strong dynamics of four-dimensional 
superconformal (SC) gauge theories plays an important role to achieve 
sfermion mass degeneracy as well as Yukawa hierarchy
\cite{KT,NS2,KNT,KNNT}.
We combine this SC approach with a string-inspired mechanism
based on geometry of extra dimensions.

Before explaining our scenario, let us first sketch 
generic features of purely string-theoretical mechanisms 
for generating hierarchical Yukawa couplings.
In heterotic orbifold models, matter fields can be assigned to 
twisted closed string states, which are localized to 
different fixed points in the compactified space.
Then their Yukawa couplings behave like $y_{ij} \sim e^{-af_{ij}^2}$, 
where $f_{ij}$ is the distance between the fixed points
corresponding to $\psi_i$ and $\psi_j$ \cite{HV}.
See also Ref.~\cite{KO}.
It does not seem easy, however, to obtain realistic Yukawa matrices 
only by this type of three-point couplings.
On the other hand, 
in a model with intersecting D-branes \cite{inter-D},
matter fields $\psi_i$ and Higgs fields $H$
arise from open strings between intersecting D-branes.
Then their couplings behave as $y_{ij} \sim e^{-aA_{ijH}}$, 
where $A_{ijH}$ is the area among three intersecting points 
corresponding to $\psi_i$, $\psi_j$ and $H$.
This approach may lead to realistic Yukawa matrices, 
but explicit analysis has not been done yet.

The string-theoretical approaches as above are interesting 
because they are purely geometrical in nature.
Unfortunately, however, such approaches have a disadvantage
from the viewpoint of the SUSY flavor problem.
{}For instance, in the heterotic orbifold models, 
sfermion masses are not degenerate between different twisted sectors,
except for the special case in which
$F$-terms of the dilaton and overall moduli fields 
are the only source of SUSY breaking \cite{sugra}.
Generically, several moduli fields do contribute to SUSY breaking
and sfermion masses become non-degenerate \cite{sugra2}.

Alternatively, the origin of Yukawa hierarchy
has been studied within the framework of 
effective field theories or string-inspired models.
A well-known example is the Froggatt-Nielsen (FN) mechanism \cite{FN},
in which symmetry principle plays a role
of controlling higher-dimensional couplings.
In the most impressive version \cite{IR},
an anomalous $U(1)$ symmetry, 
which originates from string models \cite{DSW,KN}, 
is used to generate a suppression factor for the Yukawa couplings
through the Fayet-Iliopoulos $D$-term.
However, the problem in this approach,
especially in models with anomalous $U(1)$,
is that sfermion masses suffer from 
flavor-dependent $D$-term contributions \cite{anomalous},
which generically lead to large flavor violations.

Here, we take a recently-proposed approach 
based on field-theoretical dynamics,
following the spirit of Refs.~\cite{NS1,KT,NS2,KNT,KNNT}.
At the cutoff scale $\Lambda$ of the four-dimensional effective theory, 
we start from a non-hierarchical initial value 
of the Yukawa couplings, $y_{ij}(\Lambda)={\cal O}\fun{1}$.
Within the SUSY framework, 
the renormalization group (RG) flow of $y_{ij}$ can be written as
\begin{equation}
y_{ij}(\mu) = Z_{\psi_i}\fun{\mu,\Lambda} 
              Z_{\psi_j}\fun{\mu,\Lambda}
              Z_H\fun{\mu,\Lambda} y_{ij}(\Lambda) \ ,
\label{RG:yukawa}
\end{equation}
where  $Z_{\varphi}(\mu,\Lambda)$ stands for the chiral 
wave-function renormalization factor of a superfield $\varphi$
between $\Lambda$ and low-energy scale $\mu$.
Note that our $Z$ is the inverse of the usual one.
Our goal is to have the desired pattern of Yukawa matrices
by generating hierarchically small $Z_{\psi_i}(\mu,\Lambda)$ 
for the first and second families
as a result of four-dimensional gauge dynamics.

Such a drastic RG flow cannot be realized 
in weakly-coupled theories like the minimal SUSY SM 
with the gauge couplings $g_a$ ($a=1,2,3$).
In Ref.~\cite{NS1}, Nelson and Strassler have proposed 
coupling the SUSY SM sector (or its GUT-extensions) 
to the SC sector, which is a strongly-coupled gauge theory and 
is assumed to have an infrared fixed point \cite{BaZa,seiberg}.
Here we concentrate on the SC sector with a product gauge group
$\GSC=\prod_i\GSC^{(i)}$,
whose gauge couplings we denote by $g'_i$ ($i=1,2,3$);
we will omit the prime if no confusion is expected.
Each family of quarks and leptons $\psi_i$ couples to
SC sector matter fields $\Phi_i$ and $\bar \Phi_i$,
which are charged under the $i$-th SC gauge group $\GSC^{(i)}$,
through the `messenger' coupling
\begin{equation}
\lambda_{i\,}\psi_i \Phi_i \bar \Phi_i \ .
\label{mes-1}
\end{equation}
Since the messenger interactions should be invariant
under the SM-sector gauge group $\GSM$,
some SC-sector matter fields should also be charged under $\GSM$.
We will be a little bit more explicit on this point later.

Thanks to the messenger couplings (\ref{mes-1}), 
each quark and lepton $\psi_i$ gains 
a large and positive anomalous dimension $\gamma_{\psi_i}$ 
from the corresponding SC sector. Eventually,
Yukawa couplings $y_{ij}(\mu)$ to electroweak Higgs fields 
are suppressed by the powers of anomalous dimensions.
In the original Nelson-Strassler scenario, for instance,
it is assumed that 
the first and second families couple differently to the SC sectors,
so that different anomalous dimensions are generated 
for the first two families, $\gamma_{\psi_1}\neq\gamma_{\psi_2}$.
Then one can realize hierarchical Yukawa couplings
even though their initial values $y_{ij}(\Lambda)$ 
are similar at the cutoff scale.
After generating the desired hieararchy in the Yukawa couplings,
all the SC sectors are assumed to decouple 
at once at a certain intermediate scale $M_C$.
Phenomenologically, the decoupling should be `graceful' in the sense 
that there is no large threshold correction to the couplings
and that no dangerous coupling nor bound state is generated.

The SC dynamics has another remarkable aspect.
Within a pure superconformal field theory, 
soft SUSY breaking terms are exponentially suppressed
towards the SC fixed point, 
as was first shown (for SQCD and its dual) in Refs.~\cite{kkkz,lr}.
{}For an SC model coupled with the SUSY SM sector,
where the SM gaugino masses $M_a$ ($a=1,2,3$) are not suppressed,
the mass-squared matrix of each sfermion ${\widetilde\psi}_i$ 
converges, for any initial values, on \cite{KT,NS2}
\begin{equation}
m^2_{{\tilde \psi}_i{\tilde \psi}_j}(M_C)
\ \longrightarrow\ 
  \frac{\delta_{ij}}{\Gamma_{\psi_i}}
  \sum_a 4C\fun{R^{(a)}_\psi}\alpha_a\fun{M_C} M^2_a(M_C) \ ,
\label{convergence}
\end{equation}
which is one-loop suppressed and flavor diagonal 
before diagonalizing Yukawa matrices.
Here $\alpha_a \equiv g_a^2/8\pi^2$ and 
$C(R^{(a)}_\psi)$ is the quadratic Casimir coefficient.
The factor $\Gamma_{\psi_i}$ is determined 
by anomalous dimension $\gamma_{\psi_i}$; once we know 
$\gamma_{\psi_i}$ as a function of $g'_i$ and $\lambda_i$, 
we can proceed the Grassmanian expansion of $\gamma_{\psi_i}$ 
in terms of background superfields, which are SUSY extensions 
of couplings $g'_i$ and $\lambda_i$ \cite{softbeta}.
Thus, the convergence values depend on 
$\Gamma_{\psi_i} \sim \gamma_{\psi_i}$.
When the anomalous dimensions are different 
between the first and second families,
there remains slight non-degeneracy of sfermion masses,
\begin{equation}
m^2_{\tilde \psi_1}(M_C) - m^2_{\tilde \psi_2}(M_C)
 =  \left({ 1\over \Gamma_{\psi_1}} - { 1\over \Gamma_{\psi_2}} 
    \right)\sum_a 4C\fun{R^{(a)}_\psi}\alpha_a\fun{M_C}M^2_a(M_C) \ .
\label{diff-1}
\end{equation}
Note that this non-degeneracy is one-loop suppressed.
Thus, if radiative corrections due to the SM gaugino masses, 
which are of course flavor-blind, are large enough, 
these sfermion masses are sufficiently degenerate at the weak scale.
However, such radiative correction is small for slepton masses, 
especially for right-handed sleptons,
as was estimated in Refs.~\cite{KT,NS2}. 
See also Ref.~\cite{KNNT} for subtleties of the evaluation 
(\ref{convergence}) and GUT case.

It is the infrared convergence property of sfermion masses
that makes the SC approach attractive compared with
the conventional approaches based on geometry or symmetry.
In addition to the original Nelson-Strassler scenario,
there exist several ways of modification as we shall show later.
Moreover, a modified version of the scenario can be
realized in string models in a natural manner.

We first show how the present SC framework, 
the SUSY SM coupled with product-type SC sectors,
can be realized in string models.
To this end, we take a type IIB orientifold (type I) model
with $D9$-branes and $D5$-branes, where the extra six-dimensional 
space is compactified on $T^2_1\times T^2_2 \times T^2_3$ 
and further orbifolded by a discrete symmetry \cite{IIB,IIB2}.
We assume that 
the SM gauge groups $\GSM$ originate from the $D9$-branes.
Other open string states that have both ends on the $D9$-branes
are classified into three sectors, $C^{99}_m$ ($m =1,2,3$), 
which have a nonzero momentum along the $m$-th torus $T^2_m$.
Now let us assign the $i$-th family of quarks and leptons $\psi_i$ 
to the $C^{99}_i$ sector, \ie, we identify $i=m$.
In addition, we need the SC sectors.
Here we assume that the SC gauge group 
$\GSC=\prod_i\GSC^{(i)}$ originates from $D5$-branes.
There are three types of $D5$-branes, $D5_i$, wrapping around $T^2_i$,
and we assign $\GSC^{(i)}$ to the gauge theory on $D5_i$.
{}Furthermore, there is an open string sector connecting
the $D9$ and $D5_i$ branes.
Only in this sector, denoted by $C^{95_i}$,
matter fields are charged under both $\GSM$ and $\GSC^{(i)}$.
Therefore a natural candidate for the SC matter fields, 
$\Phi_i$ and $\bar \Phi_i$, is provided by the $C^{95_i}$ sector.
With this choice, the messenger couplings of the form 
(\ref{mes-1}) are automatically realized\,\footnote{
In fact, we do not need such a tight selection rule.
A discussion on the Yukawa hierarchy transfer scenario
in the presence of mixed messenger couplings,
$\lambda_{ij} \psi_i \Phi_j \bar \Phi_j$,
will be presented in a separate publication \cite{KNT2}.
} 
since it is known \cite{IIB,sugra3} that 
among the general trilinear couplings 
$\lambda^i_{\,jk} C^{99}_i C^{95_j}C^{95_k}$,
only the followings are allowed by the stringy selection rule,
\begin{equation}
\lambda^i_{\,jk} = g_{9\,}^{\phantom{i}}\delta^i_{j}\delta^i_{k} \ ,
\label{st-in-2}
\end{equation}
where $g_9$ is the four-dimensional gauge coupling on the $D9$-branes.
This completes our assignment of gauge groups and matter fields.
Observe that three complex dimensions just fit in the existence 
of three families in Nature, and more specifically, 
the product-group structure of the SC sector 
naturally arises in this setup.
A similar setup can be realized 
in a model with $D3$-branes and $D7_i$-branes.

Our string-theoretical realization of the SC framework
suggests a peculiar initial condition of the couplings
at the cutoff $\Lambda$, which we identify with the string scale.
In ten dimensions, the gauge coupling is determined 
by a vacuum expectation value (VEV) of the dilaton field, 
but our gauge groups $\GSM$ and $\GSC^{(i)}$ originate 
from different branes,
which can have different volume in the extra-dimensional space.
Actually, the four-dimensional gauge couplings 
on the $D9$-brane and the $D5_i$-brane, 
$g_9$ and $g_{5_i}$, are related by
\begin{equation}
\frac{1}{g^2_{5_i}}
 = \left(\frac{V_i}{V}\right)\frac{1}{ g^2_9} \ , \qquad\ie,\qquad
g^2_9 = \frac{g^2_{5_1}}{V_2V_3}
      = \frac{g^2_{5_2}}{V_3V_1}
      = \frac{g^2_{5_3}}{V_1V_2} \ ,
\label{st-in-1}
\end{equation}
where $V_i$ is the volume of $T^2_i$ (in the Planck unit)
and $V=V_1V_2V_3$.
{}For a given value of the $D9$-brane coupling $g_9$, the value 
of each gauge coupling $g_{5_i}$ depends on the volume factor
of the codimension-four space transverse to the $D5_i$-branes.
Therefore, 
\textit{the SC gauge couplings can be different with each other,
depending on the geometry of compactified dimensions}\/.
Moreover, if $V_3$ is large,
the first and second SC sectors naturally become strongly coupled.
Note that this relation (\ref{st-in-1}) holds for 
holomorphic gauge couplings in the effective field theory,
since these couplings are determined by (the real part of) 
VEVs of chiral superfields, the dilaton and moduli fields.

Next we study a possible origin of Yukawa hierarchy.
{}For concreteness we take the $\GSC^{(i)}$ gauge theory
to be an $SU(\Nc^{(i)})$ with $\Nf^{(i)}$ number of flavors;
$\Phi_i$ and $\bar \Phi_i$ belong to the fundamental and 
anti-fundamental representations under $SU(\Nc^{(i)})$.
We also assume that $\Phi_i$ and $\bar \Phi_i$ are 
the only charged matter fields in each SC sector.
The $\GSC^{(i)}$ gauge theory has a superconformal fixed point
if $(3/2)\Nc^{(i)}<\Nf^{(i)}<3\Nc^{(i)}$,
where $\Nf^{(i)}$ includes 
the dimensions of $\Phi_i$ and $\bar \Phi_i$ under $\GSM$.
{}For the initial condition of the couplings,
we will be more general for a moment 
than what is suggested by the above stringy realization,
to list various possibilities for generating Yukawa hierarchy.

To find the suppression factor $Z_{\psi_i}(M_C,\Lambda)$
of the Yukawa couplings $y_{ij}$ at the decoupling scale $M_C$,
let us write the RG flow of the messenger coupling $\lambda_i(\mu)$ 
above the scale $M_C$ as
\begin{equation}
\lambda_i(\mu)
 = Z_{\psi_i}\fun{\mu,\Lambda}
   Z_{\Phi_i \bar \Phi_i}\fun{\mu,\Lambda} \lambda_i(\Lambda) \ .
\label{wf-1}
\end{equation}
The factor $Z_{\Phi_i \bar \Phi_i} \equiv Z_{\Phi_i}Z_{\bar \Phi_i}$
can be evaluated 
by integrating the RG equation for the SC gauge coupling $\alpha_i$.
Equivalently,
we use the relation between the holomorphic and physical 
gauge couplings, ${\widehat \alpha}_i$ and $\alpha_i$,
in the SC sector \cite{novikov,holomorph},
\begin{eqnarray}
\frac{1}{\widehat \alpha_i} + \Nf^{(i)} \ln Z_{\Phi_i \bar \Phi_i}
 = F(\alpha_i)
\equiv \frac{1}{\alpha_i} + \Nc^{(i)} \ln \alpha_i + \cdots \ .
\label{scheme}
\end{eqnarray}
The function $F$ may depend on Yukawa couplings,
but we neglect such dependence here.
Using the exact result
$\widehat\alpha^{-1}(\mu)-\widehat\alpha^{-1}(\Lambda)
=\left(3\Nc-\Nf\right)\ln\left(\mu/\Lambda\right)$
for the holomorphic coupling,
the factor $Z_{\Phi_i \bar \Phi_i}(\mu,\Lambda)$ 
is evaluated to be
\begin{equation}
Z_{\Phi_i \bar \Phi_i}(\mu,\Lambda)
 = \frac{Z_{\Phi_i \bar \Phi_i}(\mu)}
        {Z_{\Phi_i \bar \Phi_i}(\Lambda)}
 = \left({\mu \over \Lambda}\right)^{-\gamma_*^{(i)}}
   \exp\left[{}-\frac{F\fun{\alpha(\Lambda)}
                     -F\fun{\alpha_i(\mu)}}{\Nf^{(i)}}\,
       \right] \ ,
\label{wf-2}
\end{equation}
where $\gamma_*^{(i)}\equiv(3\Nc^{(i)}-\Nf^{(i)})/\Nf^{(i)}$
is the absolute value of the anomalous dimension 
of SC matter fields at the fixed point.
Substituting this expression (\ref{wf-2}) into Eq.~(\ref{wf-1}),
we find that the suppression factors of the Yukawa couplings 
(\ref{RG:yukawa}) are given by a formula
\begin{equation}
Z_{\psi_i}(\mu,\Lambda)
 = \left({\mu \over \Lambda}\right)^{\gamma_*^{(i)}}
   \!\times
   \left[
   \frac{\lambda_i(\mu)}{\lambda_i(\Lambda)}
   \right]
   \times
   \exp\left[\,
       \frac{F\fun{\alpha_i(\Lambda)}-F\fun{\alpha_i(\mu)}}
            {\Nf^{(i)}}\,
       \right] \ .
\label{master}
\end{equation}
In this formula,
we can replace $\lambda_i(\mu)$ and $\alpha_i(\mu)$
with their fixed point values, $\lambda_*^{(i)}$ and $\alpha_*^{(i)}$,
unless the initial couplings are so small that
the fixed point is not reached for $\mu\ge M_C$.

The equation (\ref{master}), together with Eq.~(\ref{RG:yukawa}),
is a master formula for Yukawa hierarchy in the present SC approach, 
\ie, 
by coupling the SM sector to the SC sector with product gauge group.
Corresponding to three factors on the right-hand side 
of this formula, there are three possibilities 
to achieve the hierarchical Yukawa matrices
(provided that all the SC sectors decouple at the same scale $M_C$).
The first possibility is the Nelson-Strassler scenario, in which 
all the couplings at the cutoff scale $\Lambda$ have no hierarchy, 
but the SC-sector gauge theories have 
family-dependent fixed points and anomalous dimensions.
Thus, the Yukawa couplings $y_{ij}(M_C)$ are suppressed
in a flavor-dependent manner as
\begin{equation}
Z_{\psi_i}(M_C,\Lambda)
\ \sim\ \left({M_C \over \Lambda}\right)^{\gamma_*^{(i)}} \ .
\end{equation}
In this case,
the origin of Yukawa hierarchy is purely dynamical.
As we mentioned before, however, we have non-degeneracy (\ref{diff-1})
of sfermion masses although it is one-loop suppressed.

The second possibility, a scenario of {\it Yukawa hierarchy transfer},
was proposed in Ref.~\cite{KNT}.
In order to realize sufficient degeneracy of sfermion masses,
the same structure of SC sectors was assumed there;
the same gauge group, the same field content 
and thus the same fixed point.
The origin of Yukawa hierarchy was attributed to
hierarchical initial values of the messenger couplings,
$\lambda_1(\Lambda) > \lambda_2(\Lambda) \gg \lambda_3(\Lambda)$,
which are (inversely) transferred by family-independent SC dynamics 
to the desired hierarchy of $y_{ij}(M_C)$, according to
\begin{equation}
Z_{\psi_i}(M_C,\Lambda)
\ \sim\ \left({M_C \over \Lambda}\right)^{\gamma_*}
        {\lambda_* \over \lambda_i(\Lambda)} \ .
\label{y-transfer}
\end{equation}
{}Furthermore, the assumed initial hierarchy can be realized,
without spoiling sfermion mass degeneracy,
by using the FN mechanism in SC sector.
[As was explained in Ref.~\cite{KNT},
there is no room for the FN mechanism in the SM sector.]
Thus, this scenario is a hybrid of the SC approach and 
the conventional mechanism based on symmetry principle.

Now, we point out the third possibility for the Yukawa hierarchy,
which can be most naturally realized
in the string-theoretical setup described before. The idea is 
to combine a dynamical mechanism based on SC gauge theories
with a mechanism based on geometry in extra dimensions.
Suppose that as in the second scenario,
each SC sector has the same gauge group 
and the family-independent fixed point, 
$\alpha_*^{(i)} = \alpha_*$ and $\lambda_*^{(i)} = \lambda_*$.
In addition, we apply the stringy initial condition; this time, 
the initial condition (\ref{st-in-2}) of the messenger couplings
is not hierarchical, $\lambda_i(\Lambda)=g_9$, 
but the initial values of gauge couplings $\widehat\alpha_i(\Lambda)$ 
can be different.
Indeed, in our string-theoretical realization,
we have,
by using the relation (\ref{scheme}) 
with $Z_{\Phi\bar\Phi}(\Lambda)=1$
as well as the relation (\ref{st-in-1}),
\begin{eqnarray}
{}F\fun{\alpha_i(\Lambda)}
 = \frac{1}{\widehat\alpha_i(\Lambda)}
 = \frac{8\pi^2}{g^2_{5_i}}
 = \left(\frac{V_i}{V}\right)\frac{8\pi^2}{g^2_9} \ ,
\label{matching}
\end{eqnarray}
where in the second equality,
we have made the tree-level matching of gauge couplings 
between the effective and string theories.
Substituting this into the master formula (\ref{master}),
we finally arrive at
\begin{equation}
\varepsilon_i \equiv
Z_{\psi_i}(M_C,\Lambda)
 = C \exp\!\left[\,\frac{8\pi^2}{\Nf\,g^2_{5_i}}\,\right]
 = C \exp\!\left[\,\frac{8\pi^2}{\Nf\,g^2_9}
                   \left(\frac{V_i}{V}\right)\right] \ ,
\label{transfer:geometry}
\end{equation}
where $C$ is a universal suppression factor 
\begin{equation}
C \equiv \left({M_C \over \Lambda}\right)^{\gamma_*}
         \left(\frac{\lambda_*}{g_9}\right)
         \exp\!\left[{}-{F\!\left(\alpha_*\right) \over \Nf}\,
               \right] \ .
\end{equation}
Thus, the wave-function suppression factor 
$\varepsilon_i=Z_{\psi_i}(M_C,\Lambda)$ 
for the $i$-th family of quarks and leptons
depends on geometric data of the compactification, \ie,
the volume $V_i$ of the $i$-th torus.
As the volume $V/V_i$ of the transverse space 
of the $i$-th $D5$-brane becomes large,
the initial value of the $\GSC^{(i)}$ gauge coupling becomes large 
and the factor $\varepsilon_i$ becomes less suppressed.

In this way, the (holomorphic) gauge couplings in SC sectors 
can be different at the string scale $\Lambda$ 
if the compactified extra-dimensional space is anisotropic,
and such difference is again transferred 
by flavor-independent superconformal dynamics
to the desired hierarchy of Yukawa couplings of quarks and leptons.
{}For example, the Cabbibo angle can be explained
if the size of the first and second tori is slightly different
in such a way that
\begin{equation}
\varepsilon' \equiv \frac{\varepsilon_1}{\varepsilon_2}
 = \exp\!\left[\,\frac{8\pi^2}{\Nf\,g^2_9}
               \left(\frac{V_1-V_2}{V}\right)\right]
\sim 0.22 \ .
\end{equation}
Moreover, if only the volume $V_3$ of the third torus 
is large (in the Planck unit),
\begin{equation}
V_3 \,\gg\, V_2 \,>\, V_1 \,\sim\, 1 \ ,
\end{equation}
the corresponding gauge coupling $g'_{3}\sim g_{5_3}$ is as small 
as the SM-sector gauge couplings $g_a\sim g_9$
and will not reach the fixed point.
Then the $C^{99}_3$ sector of quarks and leptons do not receive
much suppression, leaving the top Yukawa coupling of order one.

As an illustration, we give a toy model with 
$\GSM=SU(3)_C$ and three families of quarks.
Now the SC matter fields $\Phi_i$ and $\bar \Phi_i$ belong to 
$(\bm{3}+\bm{\wb{3}},\bm{\Nc})$ and 
$(\bm{3}+\bm{\wb{3}},\bm{\wb{N}}_{\!{\rm\bf c}})$ 
under $\GSM \times \GSC^{(i)}$ so that the messenger interactions 
(\ref{mes-1}) are trivially gauge invariant.
We assign the up sector of quarks $u_{\L,\R i}$ to 
the $C^{99}_i$ sector, respectively for $i=1,2,3$.
If we assume that $V_1 < V_2 \ll V_3$,
then the up-sector Yukawa matix at the scale $M_C$ takes the form
\begin{eqnarray}
y_{u ij} &\sim & \pmatrix{
\varepsilon_1^2 & \varepsilon_1 \varepsilon_2 & \varepsilon_1 \cr 
\varepsilon_1 \varepsilon_2 & \varepsilon_2^2 & \varepsilon_2 \cr 
\varepsilon_1 & \varepsilon_2 & 1 } \ .
\end{eqnarray}
As for the down sector of quarks,
we assign the left-handed quarks $d_{\L i}$ as above.
A realistic Yukawa matrix can be obtained
by assigning their right-handed quarks 
$d_{\R 1}$, $d_{\R 2}$ and $d_{\R 3}$ to 
$C^{99}_1$, $C^{99}_2$ and $C^{99}_2$, respectively,
\begin{eqnarray}
y_{d ij} &\sim & \varepsilon_2
\pmatrix{
\varepsilon' \varepsilon_1  & \varepsilon_1 & \varepsilon_1 \cr 
\varepsilon' \varepsilon_2 & \varepsilon_2 & \varepsilon_2 \cr 
\varepsilon' & 1 & 1 } \ .
\end{eqnarray}
Here we have taken the factor $C$ common,
but in principle these can be independent 
for left and right, and/or up and down sectors.
In this way, 
realistic Yukawa matrices can be obtained
even in this simple toy model 
(as far as quark sector is concerned).

We have not taken into account the stringy selection rule 
among $C^{99}_iC^{99}_jC^{99}_k$ couplings.
Unfortunately, the stringy selection rule allows 
only the coupling with $(i,j,k)=(1,2,3)$ and its permutations.
Therefore, if the electroweak Higgs fields 
are assigned to one of $C^{99}_i$ sectors, 
some of necessary Yukawa couplings are not allowed.
A possible wayout is to assume that the light Higgs fields 
would be linear combinations of Higgs fields from several sectors.
Alternatively, one could suppose that for some reasons,
effective field theory does not respect 
all of the stringy selection rules.
As a matter of fact, 
in most of string models like those sketched at the beginning,
the selection rules are too restrictive for three-point couplings 
as well as higher-point couplings \cite{higher}.
It remains a challenging problem
to derive realistic Yukawa couplings from string theory,
especially with the minimal number of Higgs fields.

{}Finally we compare our string-inspired scenario 
with the purely geometrical approaches.
The main difference lies in the sfermion mass spectrum.
At the cutoff scale $\Lambda$, we have generically non-universal 
sfermion masses among three $C^{99}_i$ sectors
to which three families of quarks and leptons belong.
{}For example, when the SUSY is broken 
by $F$-components of dilaton and moduli fields,
soft scalar masses can be parametrized as \cite{sugra3}
\begin{eqnarray}
m^2_{C^{99}_i}
 &=& m_{3/2}^2\left(1-3\Theta_i^2\cos^2\theta\right) \ , 
\end{eqnarray}
where $m_{3/2}$ is the gravitino mass and 
$\theta$ and $\Theta_i$ are goldstino angles.
In the conventional approaches, sfermion masses at the weak scale 
are also non-degenerate for generic parameter space.
In our scenario of string-inspired hierarchy transfer, 
on the other hand,
the family-independent structure of SC dynamics guarantees 
the denegeracy below the decoupling scale $M_C$
of the superconformal field theories.

To summarize, we have proposed 
an extra-dimensional scenario that simultaneously leads to 
hierachical Yukawa couplings and degenerate sfermion masses.
Our scenario combines 
the conventional geometrical approach to Yukawa hierarchy with 
a new dynamical approach based on superconformal field theories,
and can have a natural realization in string models\rlap.\,\footnote{
Some examples of an explicit string model which contains a subsector
within conformal window were constructed in Refs.~\cite{IIB,IIB2}.
}
The volume suppression in extra-dimensional spaces can generate 
the difference among initial values of SC gauge couplings,
and such difference is transferred by the superconformal dynamics 
to the hierarchy of Yukawa couplings of quarks and leptons.
The same dynamics makes sfermion masses degenerate.

In this letter, 
we have confined ourselves to the SC sector of product-type
and assumed their simultaneous decoupling from the SM sector.
A concrete mechanism for the decoupling as well as 
the `unification' of SC sectors into a simple group 
will be discussed elsewhere.

\section*{Acknowledgments}
The authors would like to thank Yoshihisa~Yamada for discussions.
The work of T.\ K.\ is supported in part 
by Grant-in-Aid for Scientific Research
from Ministry of Education, Science, Sports and Culture of Japan
(\#14540256).
The work of T.\ N.\ is supported in part by the Japan Society 
for the Promotion of Science under the Predoctoral Research Program.


\end{document}